\documentclass[11pt]{article}
\usepackage[utf8]{inputenc}
\usepackage[T1]{fontenc}
\usepackage{graphicx}
\usepackage{grffile}
\usepackage{longtable}
\usepackage{wrapfig}
\usepackage{rotating}
\usepackage[normalem]{ulem}
\usepackage{amsmath}
\usepackage{textcomp}
\usepackage{amssymb}
\usepackage{capt-of}
\usepackage{hyperref}
\usepackage[spanish]{babel}

\author{W. Luis Mochán}
\date{\today}
\title{Entrelazados}
\hypersetup{
 pdfauthor={W. Luis Mochán},
 pdftitle={Entrelazados},
 pdfkeywords={},
 pdfsubject={},
 pdfcreator={Emacs 27.1 (Org mode 9.3)},
 pdflang={Spanish}}
\begin{document}

\maketitle
\section{Resumen}
Un pequeño diálogo sobre entrelazamiento cuántico.
\section{Introducción}
\label{sec:org6a57598}
En 2019 fui invitado a participar en una mesa redonda titulada
\emph{Entrelazamiento cuántico, Alicia en el país de la física y la ficción de lo
extraño}, como parte de un evento universitario sobre literatura. La
\emph{mecánica cuántica} es una teoría extraña, como extrañas son las
aventuras de \emph{Alicia en el país de las maravillas}, pero muchos de
los aspectos raros de la mecánica cuántica se vuelven relativamente
comprensibles en el contexto de la física ondulatoria, esa que
explica la propagación del sonido, la luz, los sismos y las
olas del mar.  Sin embargo, hay un aspecto de la mecánica cuántica,
el \emph{entrelazamiento o enredamiento cuántico}, que es mucho más extraño
y que no tiene analogía en la física
clásica de las ondas.   Entre otras cosas,
viola los principios de ser una teoría \emph{local y realista}. Para
ilustrar qué significa esto, y mostrar la magia del entrelazamiento. y
siguiendo el ejemplo de un colega [ver
referencia 1] es que elaboré el siguiente
diálogo. Debo hacer notar que el tema del diálogo puede parecer
banal y poco interesante. Para entender su fondo será necesario para
el lector estar muy despierto y atento a detalles aparentemente
insignificantes, y probablemente deba tomar una hoja en blanco y un
lápiz y hacer algunas cuentas simples.

\section{Entrelazados}
\label{sec:orgf43e104}
\begin{itemize}
\item ¡No puede ser! ¡Es imposible!
\item ¿Qué te pasa Beto?
\item Esto es incompatible con cualquier definición razonable de realidad.
\item Seguro exageras.
\item No Carlos, te lo describiré, pero es todo muy sutil y te tienes que
concentrar para entender el asunto que me preocupa.
\item Soy todo oídos.
\item Tú sabes que soy obsesivo.
\item Y compulsivo, diría.
\item Y que llevo una bitácora en la que anoto \emph{todo}, hasta los detalles
más irrelevantes.
\item No me sorprende.
\item Pues Alicia también.
\item Siendo tu hermana, es natural que comparta tus obsesiones.
\item Ya sabes que en mi familia somos muy poco comunicativos, y desde que
se fue del Valle no había hablado con ella, pero ayer le
llamé.
\item Ya era hora. Y ¿qué cuenta?
\item Tristeaba con nostalgia cuando nos dejó y extrañaba la comida
del pueblo, hasta que descubrió tres fondas: \emph{La Flaca}, \emph{Don Gordo} y
\emph{Los Huesos}.
\item ¡Qué casualidad! Así se llaman mis fondas favoritas.
\item Y las mías. Curiosamente, las tres tienen sucursales aquí y en la
Montaña y sirven los mismos platillos. Resulta que, al
igual que yo, Alicia come cada día en una de ellas y todos los
días pide una comida corrida.
\item ¡Seguro que ella debe tener algo más relevante que contarte!
\item Desde luego, pero esta historia se vuelve \emph{muy} interesante, aunque
tendrás que escucharla completa para apreciarlo.
\item Bien, continúa.
\item Como yo, Alicia escoge \emph{al azar} a qué fonda ir cada día, tirando
un dado. Ayer le tocó comer en Don Gordo
y a mí también. El menú era el de siempre y escogimos exactamente los
mismos platillos. Al final, por ejemplo, nos ofrecieron café o postre y ambos
pedimos \emph{café}.
\item No es una gran coincidencia.
\item Por hacer conversación, le pregunté por el día
anterior. Anteayer ambos comimos en La Flaca y los dos pedimos
\emph{postre}. Intrigado, saqué mi bitácora y ella la suya. Empezamos a
comparar todas nuestras comidas del último año.
\item ¡Lo que es no tener qué hacer ni de qué hablar! Y ¿qué hallaron?
\item Que nuestra elección de fonda fue efectivamente \emph{aleatoria}. En el año
visitamos alrededor de 120 veces cada una de las tres, sin ningún orden en
particular. Terminamos cada comida con café o postre
indistintamente, unas 180 veces cada elección, de nuevo, sin orden
alguno y sin importar en qué fonda estábamos.
\item No es muy sorprendente.
\item Claro, aún no llego a lo extraño.  Coincidimos en fondas equivalentes
en la tercera parte de las ocasiones: En alrededor de
cuarenta ocasiones ambos comimos en La Flaca, en otras cuarenta
ambos fuimos a Don Gordo y en similar número de días ambos
terminamos en Los Huesos.
\item Eso es lo esperado \emph{estadísticamente}, la elección es al azar, el año tiene
poco más de trescientos sesenta días y hay nueve posibles parejas de
fondas, 360=9 \texttimes{} 40.
\item Pero \emph{cada una de las veces que coincidió nuestra
elección de fonda ¡elegimos los mismos platillos!} Por ejemplo, en
las más o menos veinte veces en que ambos comimos en Don Gordo y
en que, además, ella pidió café, yo también pedí café; en las veinte
ocasiones en que
ambos comimos en Don Gordo y ella pidió postre, yo también pedí
postre. Además, en las veinte ocasiones que ambos fuimos a La Flaca
y ella pidió café, yo también pedí café y en las veinte ocasiones\ldots{}
\item Ya, ya, ya entendí. Aquellas veces que eligieron la  misma fonda
escogieron los mismos platillos. Eso sí es algo raro. Yo hubiera esperado
que si ambos son indiferentes ante la elección de fonda y les da igual
pedir café que postre, la mitad de las veces en que ella pidió
café tú hubieras pedido café y la otra mitad de las veces tú
hubieras pedido postre, sin importar en qué fonda estaba cada
uno. Pero por otro lado, son hermanos. Quizás haya algo en sus genes
o en su educación que los predispone a pedir el mismo platillo
cada día. Como si cada mañana se levantaran ambos y recibieran un
mensaje común que los incite a ambos bien a tomar café o a tomar postre.
\item Todas las mañanas ambos recibimos un pequeño mensaje del Viejo quien
rutinariamente nos escribe \emph{'buenos días'}, pero nada más, y nunca le
contestamos. Pero la explicación tiene que ser más complicada,
pues \emph{no pedimos lo mismo siempre; sólo cuando ambos vamos a la
misma fonda,} y
no sabemos a qué fonda iremos sino hasta el último momento en que
tiramos el dado.
\item Entonces tendrían que amanecer cada mañana programados de alguna
manera, con instrucciones más detalladas del
tipo: si vas a La Flaca (F) pedirás café (C), si vas a Don Gordo (G)
pedirás postre (P) y si vas a Los Huesos (H) pedirás café (C).
\item Eso pensé en un principio. Si de alguna manera yo amanezco con un
juego de instrucciones,
digamos C, P y C, correspondientes a las fondas F, G y H, y ella
amaneciera con las mismas instrucciones CPC,
ambos pediríamos lo mismo en caso de ir al mismo changarro pero
podríamos pedir algo distinto para ciertos pares de fondas.
\item Lo curioso es que \emph{nunca podrías averiguar} el juego de
instrucciones completo.
\item Así es. Sólo averiguaría \emph{las instrucciones correspondientes a la
fonda a la que sí fui, y a la que fue Alicia}, de ser distinta. Nunca
podría averiguar la instrucción restante.
\item Eso es más curioso entonces. Si cada quien recibe mágicamente un
juego de tres instrucciones, pero sólo puede averiguar una, o a lo
más, dos de ellas, la restante existe, pero no es observable. Es un bonito
problema para un filósofo. ¿En qué sentido existe algo que no es
observable?
\item Me recuerda al problema de la existencia de los números
innombrables, incalculables e indescriptibles. Pero mi hallazgo es
más sorprendente. Resulta que a lo largo del año pedimos lo mismo
aproximadamente la mitad de las veces.
\item ¿Y eso qué tiene de raro?
\item De haber un juego de instrucciones, estarás de acuerdo que cada día
tendría que ser uno de los siguientes ocho: CCC, CCP, CPC, CPP, PCC,
PCP, PPC, o PPP, y no habría más (ver figura 1).
\begin{figure}[htbp]
\centering
\includegraphics[width=.9\linewidth]{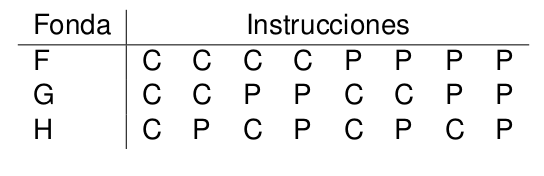}
\caption{\label{fig:org2ed83da}Todos los posibles juegos de instrucciones indicando a cada hermano qué pedir (café (C) o postre (P)) en cada restaurante (La Flaca (F), Don Gordo (G) o Los Huesos (H)).}
\end{figure}
\item De acuerdo.
\item Si cierto día recibiéramos las instrucciones CCP. Ambos
coincidiríamos en pedir café en los casos FF, FG, GF, GG, ambos pediríamos postre en el
caso HH, y discreparíamos en los cuatro casos restantes FH, GH,
HF y HG. Para los otros juegos de instrucciones pasaría lo mismo,
coincidiríamos para cinco y discreparíamos para cuatro de las
posibles elecciones de fonda,
excepto para los casos CCC y PPP, en los cuales coincidiríamos para \emph{todas} las
parejas de fondas. Entonces, de haber instrucciones, debiamos haber
coincidido en cinco o más de cada \emph{nueve} comidas. Pero sólo lo hicimos en una de cada dos
comidas, es decir en cinco de cada \emph{diez}.
\item Caray, la diferencia entre 5/9 y 5/10 no es mucha.
\item Pero es \emph{significativa}. ¡Debía ser imposible coincidir en menos de
5 de cada 9 comidas!
\item Entonces no puede existir juego alguno de instrucciones. Tu
estadística muestra que sus acciones no estuvieron predeterminadas, que
ejerciste tu libre albedrío.
\item Pero entonces, ¿cómo podríamos explicar entonces los casos en
que sí acabamos en la misma fonda? En ellos, nuestra correlación fue
perfecta. ¿Será que sin darnos cuenta nos
comunicamos telepáticamente y nos ponemos de acuerdo?
¿O hay un complot de fondas que nos manipulan como parte de un
sofisticado juego perverso?
\item Hmmm\ldots{} Te tengo que contar que hace poco asistí a una charla de
física sobre \emph{entrelazamiento
cuántico}. Desde luego, no entendí mucho, pero se me ocurre, ¿no será
que con su mensaje diario de buenos días, tu viejo les manda cada mañana
una pareja de partículas \emph{enrtrelazadas cuánticamente} cuyo estado colapsa
cuando ordenan la comida y determina sus elecciones?
\end{itemize}

\section{Discusión}
\label{sec:orgf1df9c3}

Que si Alicia o Beto tomaron café o postre puede parecer un
tema insignificante, pero lo aquí narrado
violaría principios fundamentales como el de la existencia de
una \emph{realidad local}, y esta situación
podría realizarse si, como sugiere Carlos al final del
diálogo, el padre enviara a cada hijo una de dos partículas
entrelazadas cuánticamente como podrían ser dos \emph{electrones} o dos
\emph{fotones} (referencias [2] y [3]) y midieran alguna de sus
propiedades de acuerdo a la fonda que visitan.

\begin{figure}[htbp]
\centering
\includegraphics[width=.9\linewidth]{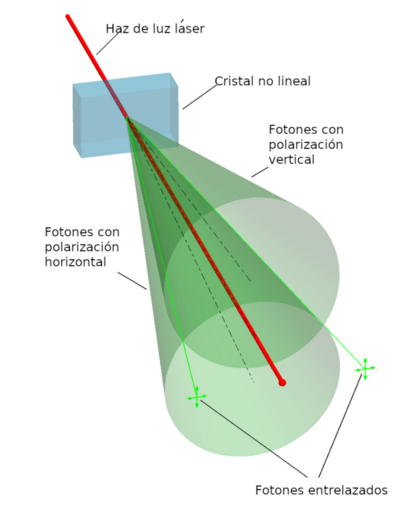}
\caption{\label{fig:orgcffb67c}Generación de dos fotones entrelazados. Un fotón linealmente polarizado (cuyo campo eléctrico oscila a lo largo de una línea recta) incide en un cristal \emph{no lineal} y se parte en dos fotones que se pueden propagar en las direcciones indicadas por la intersección de dos conos. Si se midiera la polarización de uno de ellos y resultara vertical, la del otro también lo haría. Si la del primero resultara horizontal, la del segundo también. Sin embargo, ninguno de los dos tiene una polarización real, bien definida, antes de que se mida o que se mida la de su compañero. (Wikipedia, \url{https://bit.ly/3pdUr2d}).}
\end{figure}

  En la figura 2 se muestra una forma de producir fotones
  correlacionados. La luz, como algunos lectores recordarán, está
  formada de partículas llamadas \emph{fotónes} que están asociados a un
  campo eléctrico que oscila a lo largo de cierta dirección llamada
  \emph{polarización} y que está acompañado de un campo magnético que
  oscila en la dirección perpendicular. Un solo fotón con
  \emph{polarización lineal}, se puede dividir en dos
  fotones al atravesar ciertos \emph{cristales no lineales}. Si se midiera la
  polarización de uno de ellos mediante un \emph{polarizador} (ver figura 3)
, se vería
  que como apunta en una dirección, apunta en cualquier otra. La mitad
  de las veces pasaría y la otra mitad sería absorbido por el
  polarizador, independientemente de la orientación del mismo. Sería
  un fotón \emph{no-polarizado}.

\begin{figure}[htbp]
\centering
\includegraphics[width=.9\linewidth]{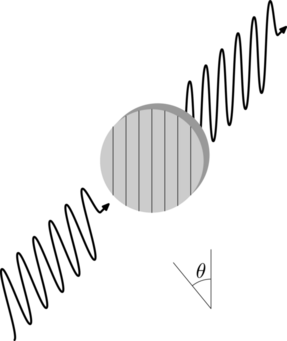}
\caption[.]{\label{fig:org5e3d7ac}Un fotón polarizado a un ángulo \(\theta\) respecto a la vertical puede atravesar un polarizador vertical con una probabilidad cos\textsuperscript{2}\(\theta\) y convirtiéndose en un fotón con polarización vertical, o puede ser absorbido con una probabilidad sin\textsuperscript{2}\(\theta\). Un fotón no polarizado pasaría a través del polarizador con probabilidad 0.5, independientemente de la orientación del polarizador.}
\end{figure}

Sin
embargo, si un fotón atraviesa un polarizador vertical, su pareja
atravesaría \emph{con certeza} otro polarizador vertical. Lo mismo pasaría
si ambos polarizadores fueran horizontales o tuvieran cualquier otra
dirección coincidente. Es como si la
polarización del segundo fotón se definiera al instante en que
medimos la polarización del primero y fuese idéntica a la que
adquiere éste. Si el padre les enviara uno de
estos fotones a Alicia y el otro a Beto, si ellos tuvieran cada uno un
polarizador de luz y lo orientaran a lo largo
de una de tres direcciones distintas, digamos, en la dirección de las 4,
las 8 o las 12 horas de un reloj de manecillas, de acuerdo al
restaurante (\emph{F}, \emph{G} o \emph{H}) elegido, y escogieran café si el fotón
atravesara su polarizador y postre si no, sus elecciones serían
consistentes con las que tomaron en el diálogo. Sus pedidos son
consistentes entonces con un proceso cuántico de decisión, pero son
inconsistentes e incomprensibles desde un punto de vista clásico.

\begin{figure}[htbp]
\centering
\includegraphics[width=.9\linewidth]{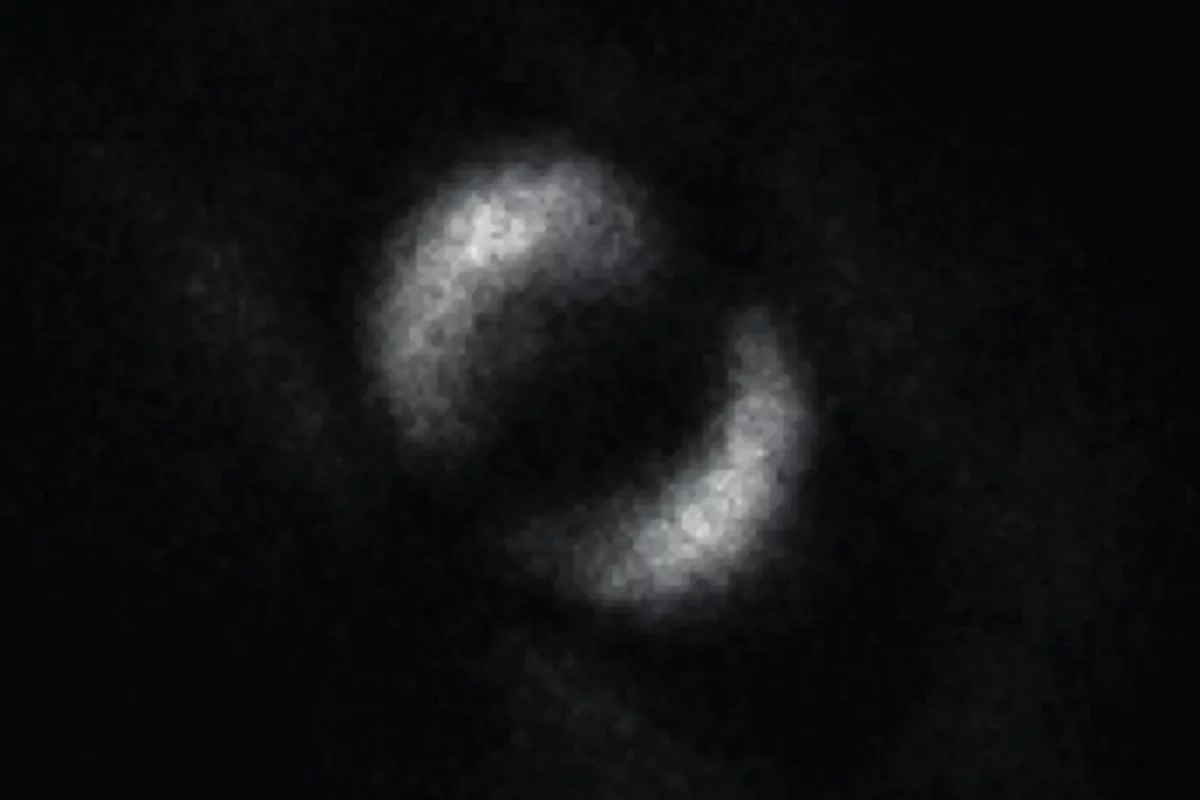}
\caption{\label{fig:org5d6248f}Una de las primeras imágenes de dos fotones entrelazados. \url{https://bit.ly/3pqCUEh}}
\end{figure}

En la figura 4 mostramos una de las primeras imágenes tomadas de
fotones entrelazados (ver ref. [4]).

Además de producir correlaciones
extrañas entre las decisiones de los dos hermanos, el
entrelazamiento o enredamiento cuántico tiene
aplicaciones potenciales que revolucionarán nuestra
tecnología en el corto y mediano plazo, como el surgimiento de la computación
cuántica [referencia 5] y su esperada disrupción del encriptamiento tal y como lo
conocemos. Hace pocos algunos años celebraba
en un congreso el anuncio, con gran bombo y
platillo, de que una computadora cuántica de media docena de cubits o bits
cuánticos, basados en una molécula de benceno había logrado mostrar que,
con una alta probabilidad, 15 podría escribirse como 3\texttimes{} 5. Sin embargo,
hace menos de dos años Google anunció que su computadora cuántica había
finalmente logrado la \emph{Supremacía Cuántica} [referencia 6]. Más modestamente, pero de
manera más inmediata, el entrelazamiento cuántico se usa hoy en día
para encriptar comunicaciones de manera segura y proteger datos
sensibles de la vista de ojos curiosos. Por otro lado, el estudio
del entrelazamiento cuántico ha impuesto fuertes
limitaciones sobre las características que puede tener una teoría
física y que, cuando nos damos el tiempo para ello, nos puede guiar a
pensar sobre conceptos tales como el significado de la realidad y la
separabilidad de los fenómenos.
Desde luego, en este pequeño artículo no pretendí explicar en qué
consiste el entrelazamiento cuántico y cómo puede conducir a tantas
aplicaciones maravillosas, pero espero haber podido transmitir lo
extraño que es el mundo cuántico y las consecuencias que puede alcanzar
en nuestra vida cotidiana.

\section{Bibliografía}
\label{sec:orgfd43c42}
\begin{enumerate}
\item \emph{Conversación sobre mecánica cuántica.} François
Alain Leyvraz Waltz, La Unión de Morelos, lunes
20 de junio de 2016. \url{https://bit.ly/3fX4PIP}
\item \emph{Is the moon there when nobody looks?}, N. D. Mermin, Physics
Today, Abril 1985 \url{https://bit.ly/2S418aY}.
\item Quantum entanblement, Wikipedia,
\url{https://bit.ly/3ielgBU}
\item \emph{Scientists capture image of quantum entanglement for the first
time}, Nick Lavars, New Atlas,
\url{https://bit.ly/34GAwzv}.
\item \emph{Computación cuántica}, Yuri Rubo y Juia Tagüeña, \emph{¿Cómo ves?} No. 67
(junio, 2004)
\url{https://bit.ly/3gq8u0P}.
\item \emph{La supremacía cuántica ha llegado de la mano de Google y China,}
\emph{pero la computación cuántica aún nos plantea estos desafíos}
\emph{titánicos}, Juan Carlos López, Xataka (Enero 2021)
\url{https://bit.ly/2SSiZ4R}.
\end{enumerate}
\section{Pies de figura}
\label{sec:org69c450f}
\begin{enumerate}
\item Todos los posibles juegos de instrucciones indicando a cada
hermano qué pedir (café (C) o postre (P)) en cada restaurante (La
Flaca (F), Don Gordo (G) o Los Huesos (H)).
\item Generación de dos fotones entrelazados. Un fotón linealmente
polarizado (cuyo campo eléctrico oscila a lo largo de una línea
recta) incide en un cristal no lineal y se parte en dos fotones
que se pueden propagar en las direcciones indicadas por la
intersección de dos conos. Si se midiera la polarización de uno
de ellos y resultara vertical, la del otro también lo haría. Si
la del primero resultara horizontal, la del segundo también. Sin
embargo, ninguno de los dos tiene una polarización real, bien
definida, antes de que se mida o que se mida la de su
compañero. (Wikipedia, \url{https://bit.ly/3pdUr2d}).
\item Un fotón polarizado a un ángulo \(\theta\) respecto a la vertical puede
atravesar un polarizador vertical con una probabilidad cos\textsuperscript{2}\(\theta\)
y convirtiéndose en un fotón con polarización vertical, o puede
ser absorbido con una probabilidad sin\textsuperscript{2}\(\theta\). Un fotón no
polarizado pasaría a través del polarizador con probabilidad 0.5,
independientemente de la orientación del polarizador.
\item Una de las primeras imágenes de dos fotones
entrelazados. \url{https://bit.ly/3pqCUEh}.
\end{enumerate}
\end{document}